\documentclass[11pt,a4paper]{article}
\usepackage{graphicx}
\usepackage{indentfirst}
\usepackage{bbm}
\usepackage{amsmath,amssymb,amsthm}
\usepackage{amssymb}
\usepackage{latexsym}
\usepackage{graphicx}
\graphicspath{ {./images/} }
\usepackage{subcaption}

\usepackage[margin=1in,a4paper]{geometry}
\usepackage{booktabs, multicol, multirow}
\usepackage{colortbl}
\usepackage{color}
\usepackage{enumerate}
\usepackage[colorlinks,citecolor=blue,urlcolor=blue]{hyperref}
\usepackage{mathtools}	
\mathtoolsset{showonlyrefs=true}

\usepackage{setspace}
\usepackage[numbers]{natbib}

\numberwithin{equation}{section}

\theoremstyle{remark}

\theoremstyle{remark}

\allowdisplaybreaks[4]

\newcounter{stepnum}

\makeatletter

\newcommand{\Rmnum}[1]{\expandafter\@slowromancap\romannumeral #1@}
\makeatother
\begin{document}

\title{Data-driven Hedging of Stock Index Options via Deep Learning}
\author{Jie Chen\thanks{Department of Systems Engineering and Engineering Management, The Chinese University of Hong Kong. Email: chenjie@se.cuhk.edu.hk.}\and Lingfei Li\thanks{Department of Systems Engineering and Engineering Management, The Chinese University of Hong Kong. Email: lfli@se.cuhk.edu.hk. Corresponding author. }}

\date{November 5, 2021}
\maketitle

\begin{abstract}
	We develop deep learning models to learn the hedge ratio for S\&P500 index options directly from options data. We compare different combinations of features and show that a feedforward neural network model with time to maturity, Black-Scholes delta and a sentiment variable (VIX for calls and index return for puts) as input features performs the best in the out-of-sample test. This model significantly outperforms the standard hedging practice that uses the Black-Scholes delta and a recent data-driven model. Our results demonstrate the importance of market sentiment for hedging efficiency, a factor previously ignored in developing hedging strategies. 
	
\end{abstract}

\section{Introduction}\label{sec:intro}
Options hedging is an important problem in financial markets. The prevailing approach to hedging first assumes a parametric stochastic model for the dynamics of the underlying asset. The model is then calibrated to observed option prices from the market, based on which various sensitivities are computed and used to hedge the risk of options.  Popular choices include local volatility models (\cite{dupire1994pricing}), stochastic volatility models (\cite{hull1987pricing}, \cite{heston1993closed}, \cite{hagan2002managing}), jump-diffusions and pure-jump processes (\cite{cont2007hedging}, \cite{kennedy2009dynamic}, \cite{li2017pure}). Despite the prevalence of the model-based approach, it is well understood that model risk can affect the hedging result significantly. 

Recently, a data-driven approach that doesn't rely on any stochastic model for the underlying asset is proposed. This approach seeks a hedge ratio that minimizes the variance of the local hedging error
\begin{align} \label{eq:hedgeerror}
	\Delta V - \delta \cdot \Delta S,
\end{align} 
where $\Delta S$ and $\Delta V$ are the change in the asset price and option price in a short period, respectively, and $\delta$ is the hedge ratio. In Hull and White \cite{hull2017optimal}, they parametrize $\delta$ in the following form:
\begin{align}\label{eq:hullmvdelta}
	\delta_{HW}  = \delta_{BS} + \frac{\nu_{BS}}{S\sqrt{\tau}}  \left( a+b\delta_{BS} + c\delta_{BS}^2   \right),
\end{align}
where $\tau$ is the time to maturity and $\delta_{BS}$, $\nu_{BS}$ are the practitioner Black-Scholes delta and vega. These practitioner sensitivities are obtained by plugging the implied volatility of the option to the formulas of the sensititives under the Black-Scholes model. It is found in \cite{hull2017optimal} that using $\delta_{HW}$ as the hedge ratio outperforms using $\delta_{BS}$ and the delta of some calibrated local volatility and stochastic volatility models. The improvement in hedging performance is particularly significant for equity index options. 

The model in \cite{hull2017optimal} is essentially a linear regression model. Nian et al. \cite{nian2018learning} propose to model the hedge ratio using a kernel function: 
\[\delta(x) = \sum_{i=1}^{m} \alpha_i k(x_i, x),  \]
where $x$ is the vector of features, $k$ is a kernel function and $x_i$ is the $i$-th observation of the features in the training data. In \cite{nian2018learning}, the authors find that using moneyness, Black-Scholes delta and time to maturity as the features and a spline kernel for hedging performs better than using $\delta_{HW}$.

In this paper, we take the data-driven approach and apply deep learning to estimate the hedge ratio for S\&P500 index options. We are motivated by the remarkable success of deep learning in many problems and for works related to financial applications see e.g., \cite{han2016deep}, \cite{weinan2017deep}, \cite{han2018solving}, \cite{sirignano2018dgm}, \cite{sirignano2019deep}, \cite{buehler2019deep}, \cite{hu2020deep}, \cite{cao2020neural} and \cite{sadhwani2021deep}. The main advantage of deep neural networks is that they can be more flexible and powerful than conventional machine learning models (like regression and kernel approximation) in capturing nonlinearity. Thus, they hold the promise of further improving the hedging performance. However, it is also well understood that normally deep learning only outperforms conventional models when the size of data is big enough (\cite{goodfellow2016deep}). This is because deep neural networks typically have tens of thousands of parameters. If the data size is too small, they can easily overfit the data and perform poorly in out-of-sample prediction. In this paper, we focus on 
the S\&P500 index options, which are one of the most liquid options in the market. For these options, there does exist a large amount of data, making deep neural networks potentially appealing. In particular, the average number of option price quotations with trading activity in one day has been growing rapidly in the past decade, from around $270$ in year 2010 to over $2000$ in year 2019.

The rest of the paper is organized as follows. We first describe our data in Section \ref{sec:data}. Section \ref{sec:deepmodel} presents the deep learning models and Section \ref{sec:empirical} shows the empirical results. The last section concludes.

\section{Data}\label{sec:data}
We obtain data for S\&P500 index options (which are European-style) from OptionMetrics in the period from January 1, 2010 to December 31, 2019 (ten years). The database provides closing bid and ask quotes for option contracts together with their hedging sensitivites such as the practitioner Black-Scholes (BS) delta (henceforth we will simply call it BS delta). We follow \cite{hull2017optimal} to filter the data. Specifically, we remove option quotes without trading or with missing information for the bid price, ask price, implied volatility, delta, gamma, vega, and theta. We also remove options with less than 14 days to maturity and those with extreme values of delta (call options with delta less than $0.05$ or greater than $0.95$ and put options with delta less than $–0.95$ or greater than $–0.05$).

After filtering, there remain about one million price quotations for calls and puts each. 
Like in \cite{hull2017optimal} and \cite{nian2018learning}, we group calls and puts into different buckets according to their BS delta defined as
\begin{align}
	&\delta_{BS}^{call}= N(d), \quad \delta_{BS}^{put}= N(d)-1,\\
	&d=\frac{1}{\sigma \sqrt{\tau}} \left[ \ln (S/K) + (r-q+\sigma^2/2) \tau \right],
\end{align}
where $r$ is the risk-free rate, $q$ is the dividend yield, $\sigma$ is the implied volatility, $S$ is the index level, $K$ is the strike price, $\tau$ is the time to maturity and $N(x)$ is the standard normal cumulative distribution function. For example, the delta bucket $a$ contains options with delta in $[a-0.05,a+0.05)$. For both calls and puts, the closer the delta is to zero, the more out-of-money the option is. Table \ref{tab:tradevolume} displays the percentage of the trading volume of each delta bucket in the overall volume. Trading tends to be concentrated on at-the-money and out-of-money options. Specifically, buckets with delta between $0.1$ and $0.6$ for calls and between $-0.5$ and $-0.1$ for puts contribute more than $95\%$ of the total trading volume.  Near-the-money calls and deep out-of-money puts are especially popular.


\begin{table}[!htbp]
	\centering
	\begin{tabular}{c|c|c|c}
		Delta bucket & Put   & Delta bucket  & Call   \\
		\hline
		-0.1           & 0.2853 & 0.1 & 0.1793 \\
		-0.2           & 0.2127 & 0.2 &0.1679 \\
		-0.3           & 0.1450  &  0.3 &0.1365 \\
		-0.4           & 0.1359 & 0.4 &0.1275 \\
		-0.5           & 0.1730  &  0.5 &0.2663 \\
		-0.6           & 0.0299 & 0.6 &0.0795 \\
		-0.7           & 0.0102 & 0.7 &0.0234 \\
		-0.8           & 0.0052 & 0.8 &0.0117 \\
		-0.9           & 0.0024 & 0.9 &0.0075 \\
		\hline
	\end{tabular}
	\caption{Percentage of trading volume for each delta bucket of S\&P 500 index options between Jan 1, 2010 and Dec 31, 2019}\label{tab:tradevolume}
\end{table}

\section{Deep Learning Models for the Hedge Ratio}\label{sec:deepmodel}
\begin{figure}
	\centering
	\begin{subfigure}[b]{0.59\textwidth}
		\centering
		\includegraphics[width=\textwidth]{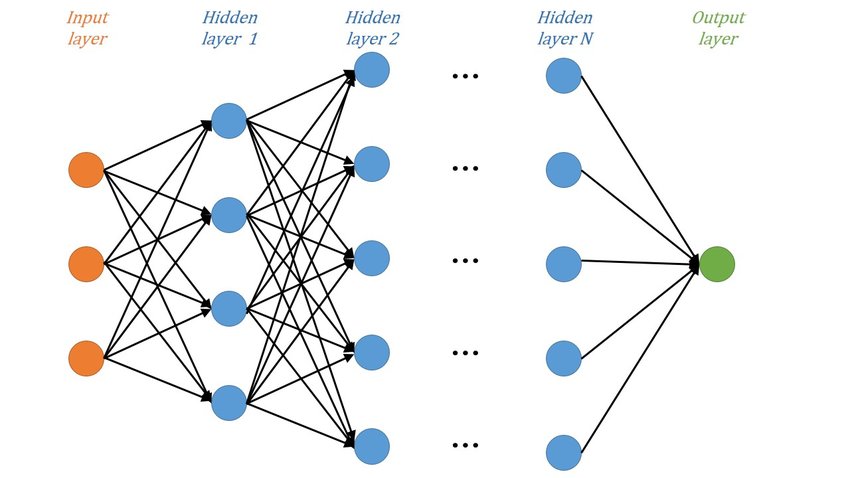}
		\caption{Structure of FNN}
		\label{fig:DNN}
	\end{subfigure}
	\hfill
	\begin{subfigure}[b]{0.4\textwidth}
		\centering
		\includegraphics[width=\textwidth]{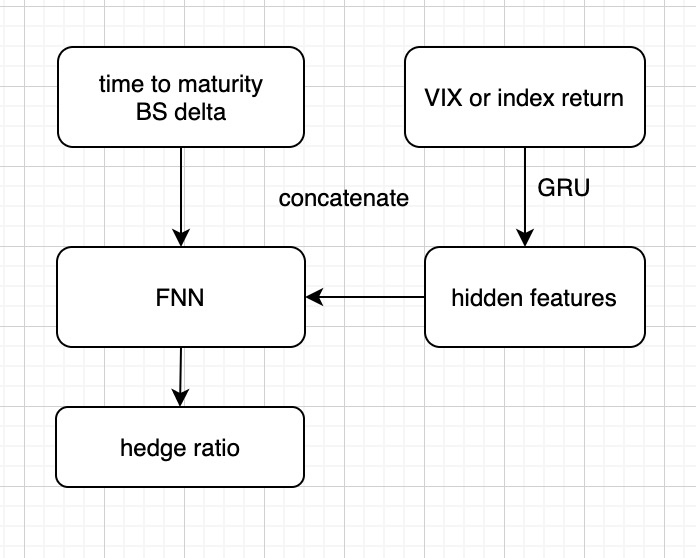}
		\caption{Structure of GRU+FNN}
		\label{fig:RNN}
	\end{subfigure}
	
	\label{fig:networkstructure}
	\caption{Structure of two types of neural network models for the hedge ratio}
\end{figure}

Let $x$ be the vector of features. We input $x$ to a feedforward neural network (FNN) and output the hedge ratio $\delta$. The structure of this neural network can be seen in Figure \ref{fig:DNN}. Suppose there are $N$ hidden layers. Mathematically, the input to output mapping is given as follows:
\begin{align}
	&h_0 = x,\\
	&h_{i+1} = f_i \left(W_i h_{i} + b_i  \right),\ \ i=0, \cdots, N-1, \\
	&\delta = \begin{cases}
		\max(W_{N} h_{N} + b_N, 0) & \ \text{for calls},\\
		\min(W_{N} h_{N} + b_N, 0) & \ \text{for puts}.\label{eq:out-activation}
	\end{cases}
\end{align} 
where $W_i$ is the weight matrix, $b_i$ is the bias vector and $f_i$ is the activation function. In our implementation, we set $N=3$ and use $128$ neurons on each hidden layer. We have also tried other values for these hyperparameters and their results are no better if not worse in out-of-sample testing.  We specify $f_i$ as ReLU (Rectified Linear Unit), i.e., $f_i(x) = \max(0, x)$, which is a popular choice for neural networks. For the output layer, we simply make the hedge ratio positive for calls and negative for puts, which are a model-free property of the hedge ratio. We can also use other activation functions for the output. For example, set $\delta=\sigma_{g}(W_{N} h_{N} + b_N)$ for calls and $\delta=\sigma_{g}(W_{N} h_{N} + b_N)-1$ for puts, where $\sigma_{g}(x)=1/(1+e^{-x})$ is the sigmoid function. However, our numerical experiment indicates that using the sigmoid function leads to slightly worse performance, which may be explained by the rather restrictive tail behavior of the sigmoid function.  

We employ two types of features as inputs to the model. One is specific characteristics of an option contract and we consider time to maturity (TTM), moneyness defined as the index level divided by the strike level, and the BS delta. Another type is concerned about market sentiment. For the hedge ratio applied to period $t+1$, we use the VIX level observed at time $t$ and the index log-return in period $t$. 
Combining these features leads to different models which are summarized in Table \ref{tab:modelfeature} (the first five ones). The full model is given by DNN3* and the others are sub-models. 

\begin{table}[!htbp]
	\centering
	\begin{tabular}{l|l}
		Model & Features  \\
		\hline 
		DNN2 & TTM, BS delta  \\
		DNN3 & TTM, BS delta, VIX for calls/index return for puts  \\
		DNN2+ & TTM, BS delta,  moneyness  \\
		DNN3+ & TTM, BS delta,  moneyness, VIX for calls/index return for puts\\
		DNN3* & TTM, BS delta,  moneyness, VIX  and index return \\
		DNN-GRU & TTM, BS delta,  VIX for calls/index return for puts\\
		\hline 
	\end{tabular}
	\caption{A summary of deep learning models for the hedge ratio}\label{tab:modelfeature}
\end{table}

Since VIX levels or index returns prior to time $t$ may also influence the hedge ratio for period $t+1$, we also consider building a sequential model for the VIX process and the index return process. To this end, we use the Gated Recurrent Unit (GRU, \cite{chung2014empirical}), which is one type of recurrent neural networks. One can also use the more complex LSTM model (\cite{hochreiter1997long}), but in our experiment doing so wouldn't bring any improvement. 

Figure \ref{fig:RNN} illustrates the structure of the new model for the hedge ratio. We first build a GRU model for VIX and another GRU model for index returns. The hidden features output by the GRU at $t$ can be viewed as summary of information up to $t$. We then input these hidden features together with other contract specific features into an FNN to get the hedge ratio for period $t+1$. This leads to the DNN-GRU model in Table \ref{tab:modelfeature}. Here we use the same features as in DNN3 because it shows the best empirical performance in our study.

The structure of the GRU cell for time $t$ is as follows. Let $\odot$ denote the Hadamard product, $x_t, h_t, \hat{h}_t, z_t, r_t$ be the input vector, hidden vector, candidate activation vector, update gate vector and reset gate vector, $W, U, b$ be the parameter matrices and vectors, $\sigma_{g}, \phi_{h}$ be the sigmoid and hyperbolic tangent function, then one GRU cell can be formulated as
\begin{align}
	& z_t = \sigma_{g} \left( W_zx_t + U_zh_{t-1} +b_z  \right),  \\
	& r_t = \sigma_{g} \left( W_rx_t + U_rh_{t-1} +b_r  \right), \\
	& \hat{h}_t = \phi_{h} \left( W_hx_t + U_h\left( r_t \odot h_{t-1}  \right) + b_h  \right), \\
	& h_t = (1-z_t) \odot h_{t-1} + z_t \odot \hat{h}_t.
\end{align}
The hidden vector $h_t$ carries the information encoded in the sequence up to $t$ and is used as one input for the FNN to generate the hedge ratio for period $t+1$.  

In our implementation, for the GRU cell, we use $8$ hidden units (i.e., $h_t$ is an eight-dimensional vector). We set the length of the sequence as $22$, the number of trading days in a month. This means $h_{22}$ only dependes on $\{x_1,\cdots,x_{22}\}$. For the FNN, we don't use hidden layers and the activation function for the output layer is the same as before. We have tried using hidden layers for the FNN, but there is no improvement in the out-of-sample test. This is because the GRU part already makes the model quite sophisticated and having a more complex FNN could cause overfitting.

\section{Empirical Study}\label{sec:empirical}

\subsection{Design of Empirical Study}
We perform daily hedging. We partition the ten years of options data into two parts: the first nine years of data for training and validation, and the last year for out-of-sample testing. We further divide the first nine years randomly into two parts:  80\% for training the model and 20\% for validation. The size of the training, validation and test set are 612770, 153193, 207283 for calls, and 878237, 219560, 299538 for puts. 

To evaluate the performance of a model $M$, we calculate its mean squared hedging error on the test set defined as
\begin{equation}
	\text{MSE}(M) = \frac{1}{N_{\text{test}}} \sum_{i=1}^{N_{\text{test}}} (\Delta V_i - \delta_i^M \cdot \Delta S_i)^2.
\end{equation}
Since the practitioner BS delta is a standard practice in the market, we follow \cite{hull2017optimal} to measure the advantage of a model $M$ over this pratice by the following gain ratio:
\begin{equation}
	\text{Gain} = 1 - \frac{\text{MSE}(M)}{\text{MSE}(\delta_{BS})}
\end{equation}

As a benchmark for our models, we estimate the Hull-White model on the training set by linear regression and report its gain on the test set. We don't consider delta hedging under popular local volatility and stochastic volatility models here because it is already shown in \cite{hull2017optimal} that using $\delta_{HW}$ outperforms these classical models. Another natural benchmark is the kernel-based model in \cite{nian2018learning}. However, estimating their model requires performing singular value decomposition (SVD) of a matrix whose size is huge with big data. The size of training data is much smaller in \cite{nian2018learning} than in our paper, and we encounter numerical difficulties to do SVD for this matrix with the size of our data even after applying the trick in their paper. For comparison on data with a moderate size so that SVD can be done efficiently, the performance of deep learning models and their kernel-based model is quite similar. This is expected as complex models like deep neural networks can only significantly outperform simpler models with sufficient amount of data.

\subsection{Training}
To train a deep learning model $M$, we minimize the loss function given by the mean squared error on the training set:
\begin{equation}
	\text{MSE}(M) = \frac{1}{N_{\text{train}}} \sum_{i=1}^{N_{\text{train}}} (\Delta V_i - \delta_i^M \cdot \Delta S_i)^2.
\end{equation}
In general, training a deep neural network is not trivial. We employ the following techniques to achieve good performance. 
\begin{itemize}
	\item \textbf{Xavier initialization and gradient clipping}: Training deep neural network faces the problem of gradient exploding or vanishing. To solve this problem, we apply Xavier initialization (\cite{glorot2010understanding}) which initializes the biases as $0$ and the weights $W_{ij}$ at each layer are sampled from the uniform distribution $U\left[-1/\sqrt{n},1/\sqrt{n}\right]$
	where $n$ is the size of the previous layer. 
	We also clip the gradient if it exceeds a threshold during backward propagation and use the clipped gradient to update the parameters.
	
	\item \textbf{Batch normalization}: This technique (\cite{ioffe2015batch}) standardizes the inputs of each layer over a mini-batch and it can stabilize and speed up the training of deep neural networks. 
	
	\item \textbf{Early stopping}: We use this technique to avoid overfitting and detailed discussions can be found in Section 7.8 of \cite{goodfellow2016deep}. While updating the neural network parameters to reduce the error on the training set, we also  monitor the error on the validation set. If the validation performance starts to degrade, we stop training.
\end{itemize}

To minimize the loss function, we use stochastic gradient descent with mini-batches of size $1024$ and the optimizer is ADAM (\cite{kingma2014adam}) with learning rate equal to $0.0005$. In our experiment, typically the training stopped after about $50$ epochs (one epoch consists of iterations needed to go through all the mini-batches).

\subsection{Empirical Findings}
We show the performance of the deep learning models and the Hull-White (HW) model (column $\delta_{HW}$) in terms of the gain for each delta bucket and the overall gain for all buckets combined in Table \ref{tab:gainSPXcall} for calls and Table \ref{tab:gainSPXput} for puts. Several conclusions can be drawn.

1. For both calls and puts, all the models show significant improvement over the standard practice which uses BS delta as can be seen from the overall gain. The best model is DNN3, whose inputs are time to maturity, BS delta and VIX for calls and index return for puts. Many models are also better than simply using BS delta for each delta bucket although some models show slightly worse performance for less liquid delta buckets. The overall gain is more significant for calls than for puts and this is also observed for the data-driven models in \cite{hull2017optimal} and \cite{nian2018learning}.

2. The DNN3 model outperforms the HW model significantly in every delta bucket and overall, and by a large margin for those actively traded delta buckets ($0.1$ to $0.6$ for calls and $-0.1$ to $-0.6$ for puts; see Table \ref{tab:tradevolume})). However, the results for other deep learning models in comparison with HW are mixed. While many of them have a greater overall gain than HW, the gain of DNN3* is less than HW for calls and the gain of DNN-GRU is less than HW for puts. Recall that DNN3* is the full model that uses all the proposed features as inputs and the DNN-GRU model considers the impact of past VIX or index returns. These results indicate that complex models are not necessarily better than simpler models even with big data if the features are not chosen properly. 

3. Comparing DNN2 with DNN2+ and DNN3 with DNN3+, one can see that with BS delta, adding moneyness as an additional input leads to inferior results. This can be explained by the relatively high correlation between BS delta and moneyness, which is $0.42$ for calls and $0.64$ for puts in our training data. In general, having highly correlated features together as inputs can create more estimation uncertainty which in turn can produce worse prediction results.

4. One also observes that DNN3 is significantly better than DNN2 and DNN3+ is significantly better than DNN2+, which reveals the importance of market sentiment for hedging. \cite{han2008investor} provides strong evidence for the existence of significant impact of sentiment on index option prices. \cite{cao2020neural} also shows that including VIX can better explain the movement of the implied volatilities of index options. However, market sentiment has not been utilized for hedging in the existing literature. To measure market sentiment, we consider VIX and index return which are easily available. It's interesting to find that having them both in the model actually increass the out-of-sample hedging error compared with having only VIX for calls and only index return for puts. One plausible explanation could be that the majority of calls and puts being traded are near the money or out of money (see Table \ref{tab:tradevolume}). Changes in the prices of these calls are very sensitive to the volatility level, while changes in the prices of these puts are very sensitive to the index return. 

5. Comparing DNN3 and DNN-GRU, we can conclude that given the current VIX level or current index return, past VIX levels and index returns are not useful. In other words, the hedge ratio depends on VIX and index return in a Markovian way.  

\begin{figure}[htbp!]
	\centering
	\begin{subfigure}[]{0.49\textwidth}
		\centering
		\includegraphics[width=\textwidth]{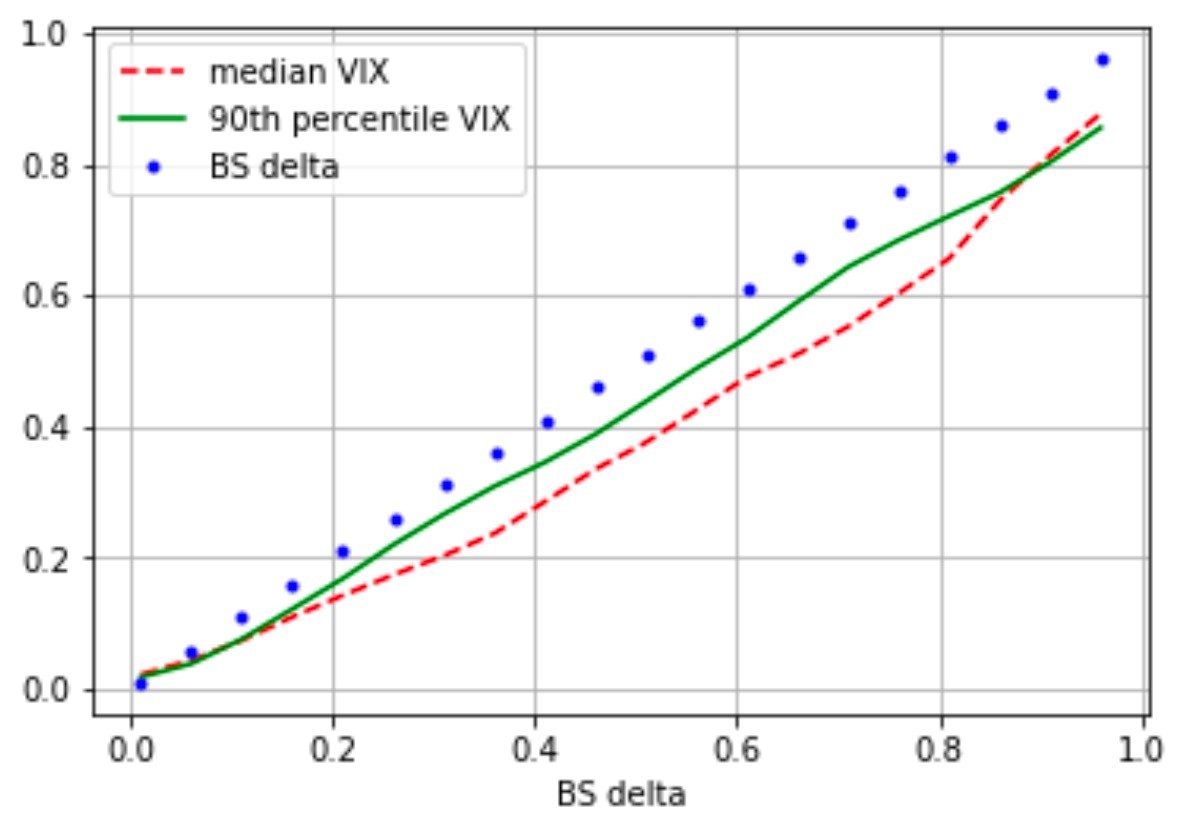}
		\caption{Predicted hedge ratio for calls with different BS delta}
		\label{fig:hedgeratiocall}
	\end{subfigure}
	\hfill
	\begin{subfigure}[]{0.49\textwidth}
		\centering
		\includegraphics[width=\textwidth]{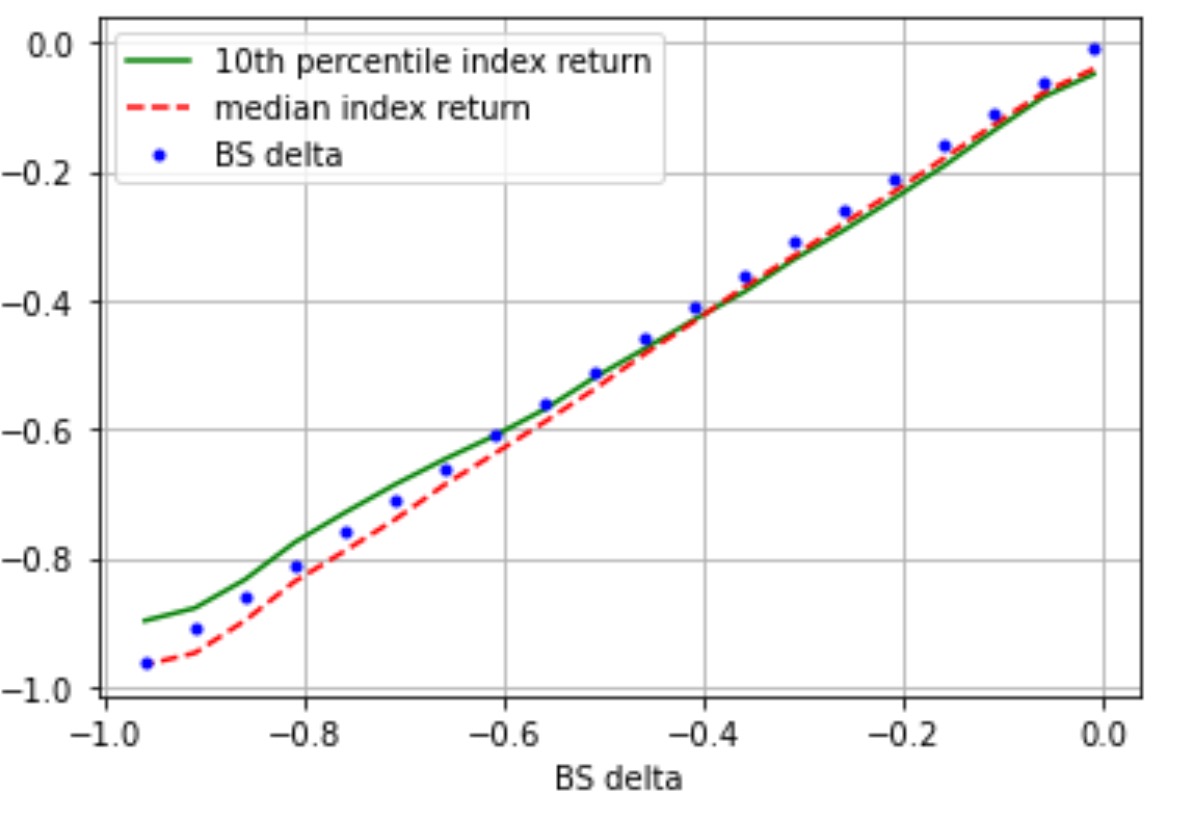}
		\caption{Predicted hedge ratio for puts with different BS delta}
		\label{fig:hedgeratioput}
	\end{subfigure}
	\caption{Predicted hedge ratio from DNN3 for options with 1 month time to maturity}\label{fig:hedgeratio}
\end{figure}

In Figure \ref{fig:hedgeratio}, we plot the predicted hedge ratio from the deep learning model DNN3 for calls and puts when the VIX or index return is at their median level in our sampling period (representing a normal market) or a level that indicates market stress. For call options, the hedge ratio of our model is smaller than the BS delta for all delta levels in both normal and stress periods. For put options, the hedge ratio of our model is greater than the BS delta in absolute values for delta in the range between $-0.5$ and $-0.1$, but the difference is much smaller compared with calls.

\begin{table}[htbp!]
	\centering
	\begin{tabular}{c|c|c|c|c|c|c|c}
		delta bucket & $\delta_{HW}$ & DNN2   & \textbf{DNN3} & DNN2+   & DNN3+  & DNN3* & DNN-GRU\\
		\hline
		0.1             & 0.1223  & 0.2082 & 0.2754 & 0.2064 & 0.2276  & 0.1047 & 0.1987\\
		0.2             & 0.1677  & 0.2036 & 0.3172 & 0.1953 & 0.2958  & 0.1486 & 0.2455\\
		0.3             & 0.1536  & 0.1842 & 0.3101 & 0.1878 & 0.2996  & 0.1577 & 0.2333\\
		0.4             & 0.1378  & 0.1853 & 0.3007 & 0.1714 & 0.2813  & 0.1474 & 0.2301\\
		0.5             & 0.1534  & 0.2110  & 0.3085 & 0.1922 & 0.2777  & 0.1635& 0.2512 \\
		0.6             & 0.2286  & 0.2699 & 0.3384 & 0.2576 & 0.2962  & 0.1795 & 0.2954\\
		0.7             & 0.2490   & 0.2857 & 0.3418 & 0.2724 & 0.2689  & 0.2020 & 0.3133 \\
		0.8             & 0.2159  & 0.2620  & 0.2639 & 0.2405 & 0.1494  & 0.1651 & 0.2972\\
		0.9             & 0.0703  & 0.1191 & 0.1037 & 0.0724 & -0.0727 & 0.0597 & 0.0907\\
		\hline
		overall         & 0.1790   & 0.2243 & 0.3077 & 0.2113 & 0.2636  & 0.1621 & 0.2580
	\end{tabular}
	\caption{Gain of various models for hedging call options. The best model is marked in bold.  }\label{tab:gainSPXcall}
\end{table}

\begin{table}[htbp!]
	\centering
	\begin{tabular}{c|c|c|c|c|c|c|c}
		delta bucket & $\delta_{HW}$ & DNN2   & \textbf{DNN3} & DNN2+   & DNN3+ & DNN3*  & DNN-GRU \\
		\hline
		-0.1             & 0.2290   & 0.1980  & 0.2785 & 0.1672 & 0.1307  & 0.0823 &0.1281 \\
		-0.2             & 0.2239  & 0.2456 & 0.3082 & 0.2006 & 0.1837  & 0.1786  & 0.2439\\
		-0.3             & 0.1972  & 0.2367 & 0.2783 & 0.2140  & 0.1936  & 0.2160 & 0.2326\\
		-0.4             & 0.1679  & 0.2079 & 0.2503 & 0.1980  & 0.2081  & 0.1979 & 0.1996 \\
		-0.5             & 0.1337  & 0.1649 & 0.2228 & 0.1907 & 0.2389  & 0.1989  & 0.1429\\
		-0.6             & 0.0918  & 0.1071 & 0.1784 & 0.1479 & 0.2087  & 0.1684  & 0.0611\\
		-0.7             & 0.0613  & 0.0672 & 0.1292 & 0.0760  & 0.1423  & 0.1160 & -0.0065 \\
		-0.8             & 0.0306  & 0.0325 & 0.0641 & 0.0144 & 0.0760   & 0.1098 & -0.0222 \\
		-0.9             & 0.0115  & 0.0341 & 0.0322 & -0.0130 & -0.0635 & -0.0019 & -0.0183  \\
		\hline
		overall         & 0.1633  & 0.1854 & 0.2403 & 0.1754 & 0.1851  & 0.1721  & 0.1607
	\end{tabular}
	\caption{Gain of various models for hedging put options. The best model is marked in bold. }
	\label{tab:gainSPXput}
\end{table}

\section{Conclusion}
This paper develops deep learning models for hedging stock index options. Our approach is data-driven, which makes no assumption on the dynamics of the underlying asset. By using a large dataset for learning, we demonstrate the advantage of deep learning over the standard practice that uses Black-Scholes delta for hedging and the data-driven model in \cite{hull2017optimal}. Among all the deep learning models considered, the best one in the out-of-sample test is  
a feedforward neural network model with time to maturity, BS delta and a sentiment variable (VIX for calls and index return for puts) as input features. Our results reveal the importance of market sentiment for hedging performance, and the best sentiment measure differs for calls and puts. 

One can also apply our deep learning models for hedging other options like individual stock options. At present, these options have far less data than S\&P500 index options, so deep learning models are unlikely to show significant advantage over simpler models. However, we can expect to see its advantage when more data becomes available in the future. 

\section*{Acknowledgements}
This research was supported Hong Kong Research Grant Council General Research Fund Grant 14206020.

\bibliographystyle{chicago}
\bibliography{NewsBib}

\end{document}